\documentclass[english,aps,preprint]{revtex4}
\usepackage[T1]{fontenc}
\usepackage[latin9]{inputenc}
\usepackage{textcomp}
\usepackage{mathrsfs}
\usepackage{amsmath}
\usepackage{amssymb}
\usepackage{graphicx}

\makeatletter
\@ifundefined{textcolor}{}
{%
 \definecolor{BLACK}{gray}{0}
 \definecolor{WHITE}{gray}{1}
 \definecolor{RED}{rgb}{1,0,0}
 \definecolor{GREEN}{rgb}{0,1,0}
 \definecolor{BLUE}{rgb}{0,0,1}
 \definecolor{CYAN}{cmyk}{1,0,0,0}
 \definecolor{MAGENTA}{cmyk}{0,1,0,0}
 \definecolor{YELLOW}{cmyk}{0,0,1,0}
}

\makeatother

\usepackage{babel}
\begin{document}

\title{Oscillations for Equivalence Preservation and Information Retrieval
from Young Black Holes}

\author{Alexander Y. Yosifov}
\email{alexanderyyosifov@gmail.com}

\author{Lachezar G. Filipov}
\email{lfilipov@mail.space.bas.bg}

\affiliation{Space Research and Technology Institute, Bulgarian Academy of Sciences}
\begin{abstract}
We follow the prevailing view that black holes do not destroy but
rather process and release information in the form of Hawking radiation.
By making certain conservative assumptions regarding the interior
dynamics of the quantum system we suggest an outside observer could,
in principle, recover the initial quantum state $\left|\psi\right\rangle $
before the black hole has evaporated half of its entropy. In the current
framework the retention time is associated with the time scale for
a local perturbation to become effectively undetectable (scrambling
time). Also, we provide a scenario for storing the information about
an infalling matter in the near-horizon region in a layered fashion.
Later we present a generic phenomena which provides a set of boundary
conditions for breaking the trans-horizon vacuum entanglement between
in- and out-modes, and thus preserve the effective field theory after
Page time $\mathcal{O}(R^{3})$. The scenario follows from black hole
perturbation theory. We further argue the proposed Planckian-amplitude
horizon oscillations can account for the physical membrane which an
observer at future null infinity $\mathcal{L}^{+}$ measures in the
complementarity conjecture. 
\end{abstract}
\maketitle

\section{Introduction}

Hawking argued in his semiclassical calculations {[}1{]} that the
radiated to infinity out-modes were of almost thermal spectrum. Particle
dependence solely on the mass of the black hole would imply lack of
correlations with the internal degrees of freedom (limited by causality)
for an outside observer. As a result, the emitted quanta will carry
no information and large amount of entropy $S=M^{2}/M_{p}^{2}$, where
$M_{p}$ is the Planck mass. In his model the lack of unitary S-matrix
and the geodesic divergence as $r\rightarrow0$ lead to loss of information.
Ever since Hawking's proposal {[}1{]} that a pure quantum state $\rho=\left|\psi\right\rangle \left\langle \psi\right|$
evolves into a mixed one $\rho=\sum_{n=1}^{N}\rho_{n}\left|\psi_{n}\right\rangle \left\langle \psi_{n}\right|$
there has been a tremendous amount of work done to both falsify and
back-up his claims {[}2-6{]}. Although many proposals regarding the
fate of information as it falls inside a black hole have been made
a consensus has still not been reached. 

Recent developments {[}7-12{]} strongly suggest the processes of black
hole formation/evaporation are described by unitary S-matrix. In the
current paper we build upon a model which, to my knowledge, was proposed
by Page {[}13{]} and advocate a scenario which suggests steady monotonic
information release throughout the black hole's evolution. For now
this seems like a physically reasonable and less radical approach.

We present a model assuming no remnants are left behind and the system
is described by unitary S-matrix. Namely, we suggest monotonic information
release begins soon after the in-modes have crossed the horizon. By
making certain reasonable (conservative) assumptions regarding the
interior dynamics of the black hole, we suggest an outside observer
does not have to wait for half of the entropy of the hole to evaporate
in order to be able to recover the initial quantum state $\left|\psi\right\rangle $
of an ingoing matter. Further, the advocated dynamics is such that
it associates the retrieval with the scrambling time scales. Moreover,
we conjecture a method for storing the information regarding infalling
matter onto the degenerate vacuum surrounding the hole in a layered-like
manner as an alternative to the conventional storing onto the horizon.
We find relations between the alternative information storing method
and black hole complementarity which further support the notion of
a physical membrane (stretched horizon) as far as a distant observer
is concerned. The reference-based complementarity descriptions establish
a time-symmetric map between past and future null infinity. Later
in the paper we use black hole perturbation theory to derive a quasi-stable
behavior of the horizon. As a result, under plausible assumptions
regarding black hole mechanics we have found that unitary interior
dynamics can cause Planckian-amplitude horizon oscillations. The conjectured
horizon oscillations may account for the stretched horizon reported
by an observer at future null infinity. The paper is organized as
follows.

In Sec. II, by making certain assumptions, we examine a scenario for
information retrieval from a black hole which has evaporated less
than half of its mass. In Sec. III we present an alternative model
for storing information regarding the quantum state of an ingoing
matter onto the degenerate vacuum in the near-horizon region. The
model we present is identical to the one given by Susskind concerning
a relation between complexity and proper distance to the horizon.
Section IV contains a brief formulation of the firewall paradox as
given by AMPS. In Sec. V we describe the quasi-stable nature of the
horizon, and further make the case for how the conjectured oscillations
may account for the stretched horizon in complementarity.

\section{Partial information reflection}

In the current Section we put forward a model motivated by an argument
made by Page {[}13{]}, certain generic assumptions regarding the fate
of information as it crosses the horizon, and the internal unitary
dynamics of the hole to make the case that \textquotedbl{}young\textquotedbl{}
black holes act as $partially\:reflecting\:mirrors.$ Until the end
of the Section we will consider the more conservative approach that
infalling information experiences no drama as it crosses the horizon
and shortly after that gets embedded onto it. Alternative proposal
will be given in Sec. III.

We define a \textquotedbl{}young\textquotedbl{} black hole to be one
which has evaporated less than half of its coarse-grained entropy.
As a result we argue an observer outside a black hole before Page
time can, in principle, recover some information regarding infallen
perturbation.

Consider the following $gedanken$ experiment. Suppose we have matter
in a box and an observer, Alice, who can freely manipulate it. Just
as it has been argued by Preskill and Hayden {[}14{]} let us assume
that Alice has studied thoroughly the matter and possesses complete
knowledge of its state. Suppose now that Alice collapses the matter
to produce a Schwarzschild black hole given by the metric

\begin{equation}
ds^{2}=-(1-\frac{2M}{r})dt^{2}+(1-2M/r)^{-1}dr^{2}+r^{2}(d\theta^{2}+sin^{2}\theta d\phi^{2})
\end{equation}

where the singularity is at $r=0$ and the global horizon is at $r=2M$.
The black hole is described by a Hilbert space $\mathcal{H}_{BH}$
with dimensionality of

\begin{equation}
dim(\mathcal{H}_{BH})=e^{A/4}
\end{equation}

in Planck units. Following the argument we further suggest Alice has
complete knowledge not only of the initial quantum state of the newly
formed hole but of its dynamics, too. Imagine now that Alice decides
to throw inside the black hole $k$ bits whose states she has also
studied. Note that no additional perturbations have been added to
the system. By assuming that Alice performs decoding once every $k$
bits are emitted, we are interested in two questions. 
\begin{itemize}
\item When will Alice be able to recover the first of the tossed $k$ bits?
\item How long will it take Alice to recover all of the bits? 
\end{itemize}
For addressing the questions we will begin by using an argument presented
by Page {[}13{]} concerning gradual information escape in the form
of Hawking radiation and will then describe the internal black hole
dynamics. A general estimate in {[}13{]} shows that information in
young black holes may, in fact, be released adiabatically. For that
reason we consider the black hole and the Hawking cloud to be subsystems
of a composite system in random pure state. The information content
in the emitted photons would be barely noticeable in a perturbative
analysis. Straightforwardly the emission will be low for a massive
black hole. In the particular model we are interested in estimating
the information content in the early radiation of a black hole which
has evaporated much less than half of its entropy. Suppose that the
two subsystems, the black hole and the Hawking emission, are described
by Hilbert spaces of dimensions $B$ and $A$, respectively. Together
they compose a larger system $AB$. The Hilbert space dimensions will
be then given by

\begin{equation}
dim(\mathcal{H}_{whole})=AB
\end{equation}

\begin{equation}
dim(\mathcal{H}_{BH})=B
\end{equation}

\begin{equation}
dim(\mathcal{H}_{radiation})=A
\end{equation}

The average information content $I$ can be approximated to be

\begin{equation}
I_{AB}=A/2B
\end{equation}

where $A$ denotes the dimensionality of the Hilbert space of the
Hawking cloud $\mathcal{H}_{radiation}$ and $B$ denotes the dimensionality
of the subsystem of the black hole where $dim(\mathcal{H}_{BH})\sim A/4$.
The initial emission of information has been calculated {[}13{]} to
be

\begin{equation}
\frac{dI}{dt}\sim e^{-4\pi/y^{2}}
\end{equation}

where $y=M_{p}/E$ and $E$ is radiation energy.

The smaller subsystem of a larger system in pure state is expected
to be in nearly maximally mixed state. For a young black hole we expect
$B>A$. As the hole evolves, however, due to the effects of the strong
gravitational field onto the quantum vacuum, the black hole will monotonically
lose mass. Based on the assumption we made in the $gedanken$ experiment
that no matter is further tossed into the hole (except for the initial
$k$ bits where $k\ll\mathcal{H}_{BH}$), as it radiates we should
see the following inversed proportionality between the subsystems
$dim(\mathcal{H}_{BH})=-dim(\mathcal{H}_{radiation})$. As a result
in the case of a young black hole we expect to find little information
in the smaller subsystem $A$. In fact most of the information is
found in the correlations between the subsystems $A$ and $B$. Detailed
calculations have been carried out in {[}13{]}. 

Based on the provided argument and a few generic assumptions given
below we address the questions regarding information retrieval time.

The notion that black holes scramble information very rapidly rather
than destroy it is embraced throughout the paper. They are conjectured
to be the fastest scramblers in nature {[}15{]}. We define a system
to be scrambled when a subsystem, smaller than half of the larger
system, reaches maximum entanglement entropy. We take the scrambling
to be a strong form of thermalization. In the context of black hole
physics, the scrambling time $t_{s}$ is the smallest possible time
scale for localized perturbation to become scrambled, (strongly thermalized)
and hence effectively undetectable

\begin{equation}
t_{s}=R\:(log\,R/l_{p})
\end{equation}

where $R$ is the Schwarzschild radius and $l_{p}$ is the Planck
length. In other words, that is the time it takes for a state to become
maximally mixed with the degrees of freedom of a system. Preskill
and Hayden have recently made the case {[}14{]} that for a sufficiently
old black hole (after Page time), the scrambling time is of order
the retrieval time $t_{s}\sim t_{retrieval}$. The retrieval time
in this context should be regarded as the time it takes an outside
observer to recover initially tossed information from a black hole
in the form of Hawking radiation. For instance, suppose we throw a
bit of classical information into an already scrambled system. As
a result the system will be briefly taken out of its present state,
and of order $t_{s}$ later will return to its initial strongly thermalized
state. Based on the provided example one might argue the scrambling
time is the time period for which the inner dynamics of an already
scrambled system \textquotedbl{}deals\textquotedbl{} with the additionally
introduced perturbation (which has, in a way, partially unscrambled
it), and thus returns to its initial scrambled state.

Moreover we suppose infalling matter experiences no drama as it crosses
the horizon $r=2M$. Further we believe in the presence of a strong
gravitational field (like the interior of a black hole as $r\rightarrow0$)
the strong energy condition (SEC) $(T_{\mu\nu}-\frac{1}{2}Tg_{\mu\nu})X^{\mu}X^{\nu}\geq0$
may be violated. Therefore as infalling matter reaches $r=0$ the
unitary dynamics transform the coarse-grained into fine-grained entropy.
Thus information is not destroyed but rather processed. It is strongly
thermalized (scrambled), and by the repulsive gravity features of
the $r=0$ region, embedded onto the horizon. The Bekenstein-Hawking
bound $S=A/4$ is satisfied as far as an outside observer is concerned.
We argue the distribution of the scrambled information across the
horizon is an exclusively stochastic process. Hence the strongly thermalized
information is not embedded onto the horizon in a perfectly uniform
manner. Homogeneous distribution would imply that each of radiated
Hawking particles carries a bit which would be problematic. In that
case the retrieval time would be of order $log\,R$ which is less
than $R\:(log\,R/l_{p})$. Information retention in a time scale smaller
than the scrambling time, Eq.(8), would allow an outside observer
to verify quantum cloning. Moreover, it is hard to come up with a
physically reasonable scenario for storing the information onto the
horizon in a specific pattern.

As a result of Page's argument and the generic assumptions we have
put forward regarding the unitary interior dynamics of a black hole
which has evaporated less than half of its coarse-grained entropy
we argue the required time for an outside observer, Alice, to recover
the first of the initially tossed $k$ bits cannot be known $a\:priori$.
Regardless of Alice's complete knowledge about the black hole's state,
dynamics and the state of the $k$ bits, there are no measurements
she can perform in order to derive a time scale for the retrieval.
The inability of determining $t_{retrieval}$ is rooted in the completely
probabilistic nature of the distribution of fine-grained entropy across
the horizon. Also, nonuniformity is necessary for deriving a value
for $t_{retrieval}$ which obeys the no-cloning bound $t_{retrieval}$$\geq t_{s}$,
and thus preserves the linearity of quantum mechanics. As we have
already made the case, if we assume that the thermalized matter is
effectively \textquotedbl{}stretched out\textquotedbl{} throughout
the horizon (perfectly uniform distribution, as one would naively
expect) we allow quantum xeroxing to be verified. Following the same
line of reasoning and expanding the argument we arrive at the same
conclusion for the case of retrieving all of the thrown $k$ bits.
Namely, Alice cannot $a\:priori$ derive a time scale. It should be
noted there is nothing, in principle, which forbids the first $k$
bits emitted from a young black hole to be that same $k$ bits that
Alice had tossed an order of $t_{s}$ earlier. However, due to the
random nature of the involved processes (scrambling and distribution
across the horizon) the probability of that occurring is exponentially
small.

We now derive several generic relations which follow from classical
black hole mechanics and thermodynamics regarding the evolution of
the information content in the Hawking radiation as a young black
hole evaporates. Let us substitute the information content $I$ from
Eq.(6) with $\Delta t$, where $\Delta t$ is the typical retention
time. We get

\begin{equation}
\Delta t=A/2B
\end{equation}

$\vphantom{}$

One can easily derive a relation between the typical retention time
$\Delta t$ and black hole's mass/area 

\begin{equation}
\Delta t=-dim(\mathcal{H}_{radiation})
\end{equation}

\begin{equation}
dim(\mathcal{H}_{radiation})=\frac{1}{16\pi M^{2}G^{2}}
\end{equation}
where $M$ is the mass of the hole and $G$ is the Newton constant.
As the black hole evaporates, the dimensionality of the Hilbert space
describing the emitted Hawking radiation cloud $dim(\mathcal{H}_{radiation})$
grows. Hence the usual time it would take $\Delta t$ for initially
thrown $k$ bits inside the hole to be retrieved by an outside observer
(Alice) decreases. This implies the information content in the radiation
$I$ is associated with the retrieval time

\begin{equation}
\Delta t=-I
\end{equation}
However, $\Delta t$ is not unbounded. As the black hole evaporates
half of its entropy, we assume we get into the realm of the argument
given by Preskill and Hayden {[}14{]}, Figure 1. Namely, as the hole
passes its \textquotedbl{}half-way\textquotedbl{} point (Page time)
$\Delta t$ reaches its lower bound

\begin{equation}
\Delta t\sim R\:(log\,R/l_{p})
\end{equation}
as $dim(\mathcal{H}_{radiation})\,\geq\,dim(\mathcal{H}_{whole})$. 

\begin{figure}
\includegraphics[scale=0.64]{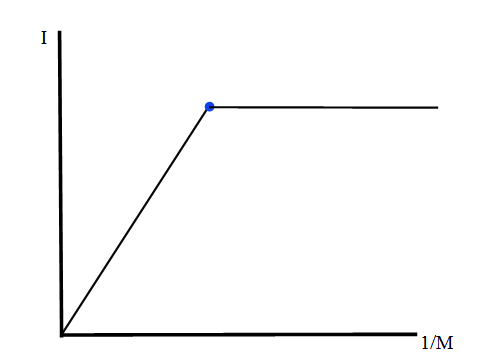}

\caption{Depiction of the information content of the emitted radiation by a
young black hole $I$ as a function of black hole's mass $M$. The
blue dot denotes the Page time at which $\Delta t\sim t_{s}.$}

\end{figure}

The provided relations Eqs. (9-12) are simple tools for more robust
illustration of the given $gedanken$ experiment and the overall model.
This further justifies our claim that black holes which have evaporated
less than half of their entropy can be regarded as $partially-reflecting\:mirrors$.
Before Page time, an observer, Alice, staying outside a black hole,
collecting and decoding radiation would expect to receive a bit every
once in a while, so to speak. We argue Alice can expect to get a complete
and rapid \textquotedbl{}reflection\textquotedbl{} of the thrown $k$
bits only after the \textquotedbl{}half-way\textquotedbl{} point has
been passed.

\section{Alternative information storing}

In this Section we consider an alternative scenario for storing information
about infalling matter. In the past, attempts have been made to falsify
the hypothesis of a complete (remnant-free) evaporation with gradual
release of information {[}21,22{]} by stating that locality has to
be violated in order for information to escape from $r<2M$. However,
it was proposed in {[}21,22{]} that one can preserve locality by assuming
no information enters the interior of the black hole. The current
Section provides such a mechanism.

It has been recently argued {[}16,17{]} that an observer outside a
black hole should see slight deviations from the unique Minkowski
vacuum. Hence implying the apparent uniqueness of the quantum vacuum
is an effective field theory. As a result the near-horizon black hole
region should be surrounded by energetically indistinguishable states.
Following the postulates of black hole complementarity {[}18{]} we
suppose the degenerate vacuum should be able to store information
about infalling matter and not get physically excited. Thus an infalling
observer will not feel anything out of the ordinary (no drama). It
has been suggested the number of the different vacua $\left|\psi_{n}\right\rangle $
is the exponential of the Bekenstein-Hawking area/entropy bound; namely

\begin{equation}
\left|\psi_{n}\right\rangle =e^{A/4}
\end{equation}

where $A$ is the area of the black hole.

Spherically symmetric solutions to Einstein field equations (EFE)
describe generic collapse with the horizon region being a flat plane
with no special dynamics. Quantum vacuum in asymptotically flat spacetime
with a boundary surface leads to a certain ambiguity (Casimir effect).
In this case the number of measured eigenstates $a_{i}^{\dagger}a_{i}$,
will be observer-dependent. Here $a_{i}^{\dagger}$ is a creation
operator, and $a_{i}$ is an annihilation operator.

Suppose we have a pair of observers, Alice and Bob, each carrying
a measuring apparatus, where Alice is close to the event horizon,
and Bob is far away. We expect Alice and Bob to disagree on the number
of the produced eigenstates in the near-horizon region

\begin{equation}
\phi=\sum_{i}(a_{i}^{\dagger}f_{i}^{*}+a{}_{i}f_{i})
\end{equation}

\begin{equation}
\phi=\sum_{i}(b_{i}^{\dagger}f_{i}^{*}+b{}_{i}f_{i})
\end{equation}
Hence they measure different number of particles, $N_{A}\neq N_{B}$. 

In {[}19{]} Susskind introduced a duality which connects the distance
from the horizon to the computational complexity in order to study
the relation between black holes and complexity. To do so he conjectured
an onion-like (layered) structure in the near-horizon region. Each
layer carries different degrees of freedom. We believe similar method
can be used for storing information onto the degenerate vacuum. Suppose
we embrace the same layered structure, Figure 2.

Let's assume the thickness of each layer is of order $l_{p}$ and
that they extend in outward direction of order $R$, where $R$ is
the Schwarzschild radius. It seems reasonable to make that assumptions
since at that scale the geometry of the spacetime is flat (Minkowski
vacuum) 

\begin{equation}
\frac{R}{l_{p}}=S_{BH}
\end{equation}

In this case the layered degenerate vacuum is sufficient to effectively
reproduce the degrees of freedom of the black hole. As a result, ingoing
matter should encode the information about its initial quantum state
onto the degenerate vacuum.
\begin{figure}
\includegraphics[scale=0.65]{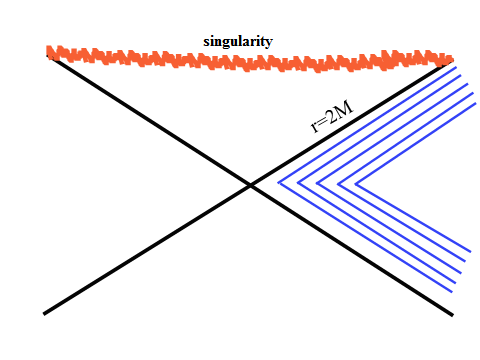}\caption{Penrose diagram for the layered (onion-like) structure of the degenerate
vacuum in the near-horizon region. The blue lines indicate the different
layers. The singularity is given by the red wave-like line.}
\end{figure}
 We further wish to extend the assumption we made about the random
embedding of scrambled information onto the event horizon. Namely,
we suggest even for the case of layered-structured degenerate vacuum
one should regard the storing of information as a completely random
process. In this case the emitted thermal Hawking photons will periodically
\textquotedbl{}pick up\textquotedbl{} information from the vacuum.
We suspect the exact mechanism for that requires insights from quantum
gravity which we do not yet possess. Thus not every single Hawking
particle will carry information. Addressing the questions raised in
Sec. II in the context of the current framework yields identical results.
This could be easily seen by the similarities between the stochastic
information encoding onto the horizon and the degenerate vacuum. Although
it may not sound as straightforward as the scenario for encoding onto
the horizon given in Sec. II, the current alternative should not be
ignored altogether as it succeeds to achieve similar results. 

\section{Postulates inconsistency}

In the present Section we examine AMPS argument regarding an inconsistency
between the postulates of complementarity.

Recently, it has been argued by AMPS {[}23{]} there is an inconsistency
between the postulates of black hole complementarity {[}18{]} which
causes drama for an infalling observer after half of the mass has
been evaporated. The three postulates are given as follows. The absence
of drama for an ingoing observer is also mentioned in {[}18{]}, and
has been established in the literature as a fourth postulate. 

$\vphantom{}$

Postulate 1: The process of formation and evaporation of a black hole,
as viewed by a distant observer, can be described entirely within
the context of standard quantum theory. In particular, there exists
a unitary S\textminus matrix which describes the evolution from infalling
matter to outgoing Hawking-like radiation.

$\vphantom{}$

Postulate 2: Outside the stretched horizon of a massive black hole,
physics can be described to good approximation by a set of semiclassical
field equations.

$\vphantom{}$

Postulate 3: To a distant observer, a black hole appears to be a quantum
system with discrete energy levels. The dimension of the subspace
of states describing a black hole of mass $M$ is the exponential
of the Bekenstein entropy $S(M)$.

$\vphantom{}$

Postulate 4: A freely falling observer experiences nothing out of
the ordinary when crossing the horizon.

$\vphantom{}$

We wish to preserve unitarity in accordance with Postulate 1

\begin{equation}
\rho=\left|\psi\right\rangle \left\langle \psi\right|
\end{equation}
Following the semiclassical approximation, stated in Postulate 2,
combined with the desired information preservation in Postulate 1,
we see that an infalling observer should encounter high-energy quanta
at the horizon, Figure 3.

However, according to Postulate 4, there should be absence of drama
for an infalling observer. Hence she should measure the ground state
with no deviations from the classical Unruh vacuum. Suppose an infalling
observer is counting the high-energy modes with a measuring apparatus.
Following Postulate 4, the expectation value should be zero, $N_{i}=0$,
where $N_{i}=a_{i}^{\dagger}a_{i}$. As AMPS have pointed out, this
statement appears to be in contradiction with our nomenclature regarding
quantum field theory in curved spacetime (Postulate 2). As Hawking
has explicitly shown in his semiclassical calculations {[}24{]} the
strong gravitational field acts on the quantum vacuum and polarizes
the virtual particle pairs. The number of the produced particles which
are radiated away to infinity is thus given by

\begin{equation}
\left\langle 0\right|a_{i}^{\dagger}a_{i}\left|0\right\rangle =\sum_{i}|\beta_{i}|^{2}
\end{equation}

\begin{figure}
\includegraphics[scale=0.64]{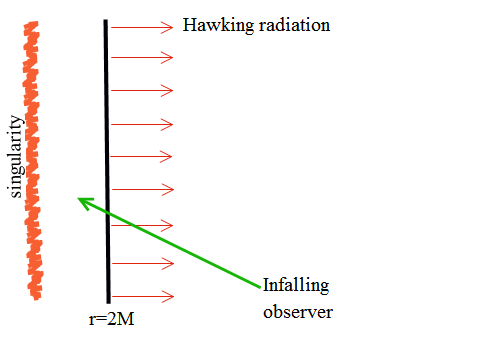}\caption{Black hole in Eddington-Finkelstein coordinates. The event horizon
is depicted by the solid black line. The solid green line depicts
an infalling observer, and the Hawking radiation is shown by the red
arrows. }
\end{figure}

By finding the value of $\beta$ we approximate the number of the
emitted quanta. As it follows from black hole perturbation theory
{[}25,26{]} as the hole evaporates it loses mass which leads to an
increase of the temperature, and thus faster evaporation rate. Although
the value of $\beta$ is generally negligible, the effects of the
black hole's mass on the matter fields add up during the course of
evaporation, and should be significant after Page time. Therefore,
after half of the black hole has evaporated the number of out-modes,
that an infalling observer carrying measuring apparatus will count,
should be non-zero, hence causing drama and contradicting Postulate
4.

Because of the contradiction between the postulates of complementarity,
AMPS argue the following three statements cannot all be true:

(i) Purity of the emitted Hawking quanta.

(ii) Absence of drama for an infalling observer.

(iii) Semiclassical physics in the vicinity of the black hole.

\section{Oscillations as stretched horizon}

We now address the question of what boundary conditions need to be
present at the vicinity of the event horizon in order to avoid formation
of a firewall for a sufficiently old black hole. We then put forward
the quasi-stable behavior of the horizon which can account for the
conjectured in complementarity stretched horizon. 

As it has been argued in Sec. IV AMPS' argument leads to violation
of the no-drama principle after Page time if there is an entanglement
(maximal correlation) between early- and late-time Hawking radiation.
It has been shown in {[}27{]} that disentangling the quantum vacuum
in the near-horizon region by introducing certain boundary conditions
preserves the effective field theory for old black holes. The imposed
boundary conditions have to meet certain requirements, namely to change
the correlation between the in- and out-modes without affecting the
thermal spectrum of the radiation emitted to $\mathscr{I^{+}}$, and
preserve the conservation of momentum. Even small deviations from
the purely thermal spectrum of the Hawking particles will lead to
stress-energy divergence due to the large blueshift which occurs when
an observer at $\mathscr{I^{+}}$traces the particles to the origin.
In that case, $T_{\mu\nu}\rightarrow\infty$ as $r\rightarrow2M$.

Following {[}18,27{]} as far as a far away observer is concerned there
is a stretched horizon (physical membrane) located $l_{P}$ away from
the global horizon ($r=2M$) in outward direction. The proposed membrane
acts as a partially-reflecting mirror with the following characteristic

\begin{equation}
\Phi_{out}-\varPhi_{in}=0
\end{equation}

where $in$ and $out$ stand for coming from $\mathscr{I^{-}}$ and
radiated to $\mathscr{I^{+}}$, respectively. Thus an observer at
$\mathscr{I^{+}}$ can obtain all of the information from $\mathscr{I^{-}}$
and vice versa. The reflective property Eq.(20) implies we preserve
the unitary evolution of the S-matrix and establish a complete time-symmetric
map between past and future null infinity. As a result the stress-energy
tensor is normalized. That being said, the conditions imposed in {[}27{]}
lead to polarization of the particle pairs solely on one side of the
horizon, either $r<2M$ or $r>2M$, and hence break the trans-horizon
correlations.

We argue the Planckian-amplitude horizon oscillations {[}20{]} can
generically account for the suggested boundary conditions. The conjectured
oscillations arise naturally from perturbation theory. The effect
is argued to be caused by transformation of coarse-grained degrees
of freedom into fine-grained degrees of freedom which is an integral
part of the black hole formation/evaporation process. We assume the
frequency of the horizon oscillations are related to the mass of the
hole following trivial principles from black hole thermodynamics.
That being said, the oscillations frequency should be gradually increasing
for a monotonically evaporating isolated black hole with no ingoing
matter present. Hence the frequency of the oscillations solely depends
on the mass of the hole

\begin{equation}
\omega=\sqrt{\frac{-T_{\mu\nu}}{M}}
\end{equation}
where $T_{\mu\nu}$ is the emitted Hawking radiation and $M$ is the
mass of the black hole. We assume the conjectured oscillations can
account for the vacuum disentanglement.

Suppose we have a spherically symmetric collapse given by the Schwarzschild
metric, Eq.(1). In the context of complementarity an observer at $\mathscr{I^{+}}$
would measure the entropy of the black hole to emerge from the fine-grained
degrees of freedom outside the global horizon (stretched horizon).
As it has been argued in {[}18{]} each point from the global horizon
($r=2M$) is projected onto a physical membrane located a $l_{P}$
away. Hence the whole horizon surface is shifted by order of $\delta$,
where $\delta$ is a small positive constant. Therefore the entropy
of the event horizon equals the entropy of the stretched horizon which
obey the Bekenstein bound in Planck units

\begin{equation}
S_{horizon}=S_{stretched}=A/4
\end{equation}
Because of the established equality we argue the oscillations can
account for the physical membrane as far as an observer at $\mathscr{I^{+}}$
is concerned. 

Since the black hole polarizes the quantum vacuum in the vicinity
of its horizon, we suppose the proposed oscillations are sufficient
to produce the desired effect, namely the particle pairs remain in
either the interior or exterior region. Thus breaking up the vacuum
entanglement. Suppose we have a collapse in initially pure state

\begin{equation}
\left|\varPsi\right\rangle =\sum_{i}\left|\psi\right\rangle \otimes\left|i\right\rangle 
\end{equation}
where $\left|\psi\right\rangle \in\mathscr{\mathcal{H}}_{out}$ and
$\left|i\right\rangle \in\mathscr{\mathcal{H}}_{in}$. Here $\mathscr{\mathcal{H}}_{out}$
and $\mathcal{H}_{in}$ stand for radiation emitted to infinity and
radiation close to the horizon, respectively. For a black hole after
Page time we assume $\left|\psi\right\rangle \;\gtrsim\;\left|i\right\rangle $.
If we interpret the radiated Hawking particles in terms of Hilbert
spaces, we get $dim(\mathscr{\mathcal{H}}_{out})\:\gtrsim\:dim(\mathscr{\mathcal{H}}_{in})$
{[}28{]}. That being said, when an observer at $\mathscr{I^{+}}$
traces the Hawking quanta back to the origin no deviations will be
observed due to the purely thermal spectrum of the emission. Moreover,
when the out-modes are traced back no membrane will be present. As
far as a close-by observer is concerned infalling matter is not reflected
by a stretched horizon, and crosses the $r=2M$ region with no drama.
We argue there will be discrepancy between the reference-based descriptions
of order the scrambling time $t_{S}$, Eq.(8), due to the lack of
perturbation to the background metric caused by ingoing matter. Hence
a close-by observer should see matter being radiated away from the
global horizon, which would imply it has been reflected by the singularity
region (dS core) {[}20{]} of order $t_{S}$ later.

So far we have provided a complementary description of the physical
membrane, and have shown how the conjectured horizon oscillations
can account for it. However, we still have not addressed the question
of what causes the infalling matter reflection, as reported by an
observer at $\mathscr{I^{+}}$. 

In Sec. III we have argued the near-horizon region is surrounded by
numerous energetically-identical vacua. Later on in Sec. V we have
shown the number of the indistinguishable vacua, Eq.(14) equals the
entropy of the global, and hence stretched horizon of the black hole,
Eq.(22). That being said, we argue a distant observer (Bob) can falsely
interpret the information stored onto the degenerate vacuum as a partially-reflective
surface. As far as Bob is concerned, infalling matter gets thermalized,
and reflected back by the stretched horizon. Moreover, the out-modes
emitted to infinity are also seen to originate from the physical membrane,
from the perspective of an observer at $\mathscr{I^{+}}$. For Alice,
however, who is at proper distance $r$ from the black hole nothing
unusual happens. Infalling matter experiences no drama, and the Hawking
particles are emitted from the global horizon ($r=2M$). 

We have been able to show that by starting from different principles
one might derive a complementary description similar to the one given
in {[}18{]}.

\section{Conclusions}

Section II was built on the premise that black hole are objects which
process information. By using mathematical framework given by Page
{[}13{]} and making certain generic assumptions regarding the unitary
interior dynamics of a black hole which has evaporated less than half
of the its coarse-grained entropy, we made the case that one does
not have to wait for Page time in order to recover information concerning
in-modes. In fact, what we have shown is that an outside observer,
having complete knowledge regarding the initial quantum state of the
black hole and the tossed matter, can, in principle, retrieve information
in the form of Hawking particles. Hence establishing young black holes
as $partially-reflecting\:mirrors$. Also, we presented several relations
which further explain the connection between the information content
in the Hawking radiation $I$ and black hole's mass $M$. As it turns
out, once the \textquotedbl{}half-way\textquotedbl{} point (Page time)
is reached the typical retention time reaches its lower bound $t_{retrieval}$$\sim t_{s}$.
Furthermore, in Sec. III we described an onion-like layered behavior
of the degenerate vacuum in the near-horizon region that can record
information about ingoing matter. In both cases, conventional (horizon)
and alternative (degenerate vacuum) we managed to obtain similar results
for the typical time scale for information retention. Thus in both
cases CPT symmetry is respected. We have then shown (Sec.V) that the
generic phenomena of Planckian-amplitude horizon oscillations can
account for the stretched horizon proposed in black hole complementarity,
and also provide the necessary conditions for breaking up the entanglement
between early- and late-time Hawking particles. Thus preventing the
formation of a firewall after Page time. The conjectured oscillations
follow from classical black hole perturbation theory. The model builds
on the complementary picture given by Susskind by providing natural
explanation of its basic features.

\end{document}